\documentclass[fleqn,twoside]{article}
\usepackage{espcrc2}

\usepackage{graphicx}
\usepackage[figuresright]{rotating}

\newcommand{\AmS}{{\protect\the\textfont2
  A\kern-.1667em\lower.5ex\hbox{M}\kern-.125emS}}

\title{Light Pseudoscalar Bosons, PVLAS and the Double Pulsar J0737-3039\thanks{Talk given by M. R. at the ``Neutrino Oscillation Workshop 2006'' (to appear in the Proceedings).}}

\author{Arnaud Dupays\address[MCSD]{INAF-IASF, Via E. Bassini 15, I-20133 Milano, Italy}   
        and 
        Marco Roncadelli\address{INFN, Via A. Bassi 6, I-27100 Pavia, Italy }}

\begin{document}

\begin{abstract}
Light Pseudoscalar Bosons (LPBs) coupled to two photons are predicted by many realistic extensions of the Standard Model and give rise to birefringence and dichroism in a light beam travelling in an external magnetic field. These effects have recently been detected by the PVLAS collaboration, thereby strongly suggesting the existence of a LPB. We provide an astrophysical cross-check for such a claim. Actually, we show that in the double pulsar J0737-3039 photon-LPB conversion can give rise to a characteristic attenuation pattern of the light beam emitted by one of the pulsars when it goes through the magnetosphere of the companion. The effect under consideration shows up in the $\gamma$-ray band and can be detected by the upcoming GLAST mission. 

\vspace{1pc}
\end{abstract}

\maketitle

\section{INTRODUCTION AND OUTLOOK}

Many realistic extensions of the Standard Model predict the existence of {\it Light Pseudoscalar Bosons} (LPBs) described by the effective lagrangian
\begin{equation}
\label{a1}
{\cal L} = \frac{1}{2} \, \partial^{\mu} \phi \, \partial_{\mu} \phi - \frac{1}{2} \, m^2 \,{\phi}^2 - 
\frac{1}{4 M} \, F^{\mu \nu} \, \tilde F_{\mu \nu} \, \phi~,
\end{equation}
where $\phi$ denotes the LPB field, $M$ is the two-photon inverse coupling constant (with the dimension of an energy), $F^{\mu \nu}$ is the usual electromagnetic field strength ($\tilde 
F_{\mu \nu}$ is its dual) and natural Lorentz-Heaviside units with $\hbar=c=1$ are adopted. A well-known example of LPB is the {\it axion}~\cite{axion}, whose mass $m$ is given by the approximate relation $m \simeq 0.7 \, ( 10^{10} \, {\rm GeV} /M)$~\cite{cgn}. However, $m$ 
and $M$ are to be regarded as independent parameters as long as generic LPBs are 
concerned~\cite{masso1}. 

Owing to the characteristic two-photon coupling in lagrangian (\ref{a1}), photon-LPB conversion takes place whenever an external magnetic field is present. This fact implies that the vacuum acquires nontrivial optical properties, much in the same way as it happens for the QED magnetized vacuum~\cite{Heisenberg}. Specifically, the exchange of {\it virtual} LPBs gives rise to {\it birefringence}, while the production of {\it real} LPBs is responsible for {\it dichroism}. Therefore, when a light beam with initial {\it linear} polarization travels in a magnetized vacuum an {\it elliptical} polarization shows up, with the ellipse's major axis {\it rotated} with respect to the initial polarization. Quite remarkably, a measurement of the resulting beam ellipticity and rotation angle permits a {\it complete} determination of $m$ and $M$, provided of course that the LPB contribution dominates over the QED one. Thus, the unique opportunity arises to detect LPBs by means of high-precision optics experiments, as first pointed out by Maiani, Petronzio and Zavattini~\cite{Maiani}.

Recently, the PVLAS collaboration has reported positive evidence for an anomalously large value of the rotation angle in an initially linearly-polarized laser beam undergoing multiple reflection in a $5 \, T$  magnetic field~\cite{pvlas}. Also the beam ellipticity has been determined on the basis of a previous experiment~\cite{cameron}. Assuming that the effect is indeed due to a LPB, the corresponding physical parameters turn out to lie in the range $ 1.0 \cdot 10^{- 3} \, {\rm eV} \leq m \leq 1.5 \cdot 10^{- 3} \, {\rm eV} $ and $2 \cdot 10^{5} \, {\rm GeV} \leq M\leq 6 \cdot 10^{5} \, {\rm GeV}$. Moreover, a new measurement of the beam ellipticity leads to the preferred values $m \simeq 1.0 \cdot 10^{- 3} \, {\rm eV}$ and $M \simeq 3.8 \cdot 10^{5} \, {\rm GeV}$~\cite{pvlas2}. As a matter of fact, preliminary evidence~\cite{pvlas3} tends to favour a scalar rather than a pseudoscalar boson, thereby requiring the last term in lagrangian (\ref{a1}) to be replaced by $(1/4M) F_{\mu \nu} F^{\mu \nu} \phi$. Yet, all considerations to follow basically hold true in both cases, and so we shall focus on LPBs for definiteness.

What is going on? A look back at the above $m-M$ relation shows that the LPB in question {\it cannot} be the axion. In addition, the inferred value of $M$ {\it violates} both the theoretical astrophysical bound~\cite{Raffelt1996} and the CAST result~\cite{cast} by about five orders of magnitudes. Such a conflict can be avoided in two ways. A possibility is that LPBs produced in the central region of a star are effectively confined inside the interiors in a manner consistent with the observed properties~\cite{masso2005}. Alternatively, the produced LPB flux can be drastically reduced in a stellar environment~\cite{masso22005}. In either case, {\it new physics} at very low energy is required. All this makes the need for {\it independent} tests of the PVLAS result even more compelling. 

Remarkably enough, high-precision astronomical observations of the double pulsar J0737-3039 can compete successfully with laboratory experiments in providing independent evidence for or against the PVLAS claim. Indeed, we will shown that photon-LPB conversion in such a system gives rise to a characteristic attenuation pattern of the light beam emitted by one of the pulsars when it goes through the magnetosphere of the companion. For the values of $m$ and $M$ determined by PVLAS, the production of LPBs turns out to be substantial, so that the resulting attenuation of the beam intensity can become observable in the $\gamma$-ray band with the upcoming GLAST 
mission~\cite{drrb}.

\section{DOUBLE-PULSAR OBSERVATIONS}

We begin by recalling a few facts about J0737-3039~\cite{burgay}. This is a double pulsar system, with an orbital period of 2.45 hr. Both components -- referred to as $A$ and $B$ -- rotate, with spin period of 23 ms and 2.77 s, respectively. What makes J0737-3039 particularly well suited for our purposes is the high inclination of its orbital plane, so that it is seen almost edge-on. 

Consider now the light beam emitted by pulsar $A$, and denote by $\rho$ its impact parameter, namely its projected distance from pulsar $B$. Every 2.45 hr $\rho$ attains its minimum value 
$\rho_0$ and the light beam in question traverses the magnetosphere of pulsar $B$. When this happens, the beam propagation is strongly affected by such a highly nontrivial environment. Although the current value of $\rho_0$ is somewhat uncertain, an estimate~\cite{coles} yields 
$\rho_0 \simeq 4 \cdot 10^3 \, {\rm km}$ and this is the value used in our analysis.

We compute the probability for photon-LPB conversion $P(\gamma \to \phi)$ by numerically integrating the propagation equation~\cite{Raffelt and Stodolsky} for a beam emitted by pulsar 
$A$ and travelling in the dipolar magnetic field of pulsar $B$ (plasma effects turn out to be totally negligible).

We plot $P(\gamma \to \phi)$ versus photon energy $\omega$ in Fig. 1, for the beam impact parameter $\rho_0 = 4 \cdot 10^3 \, {\rm km}$.

\begin{figure}[h]
\begin{center}
\includegraphics[width=8cm]{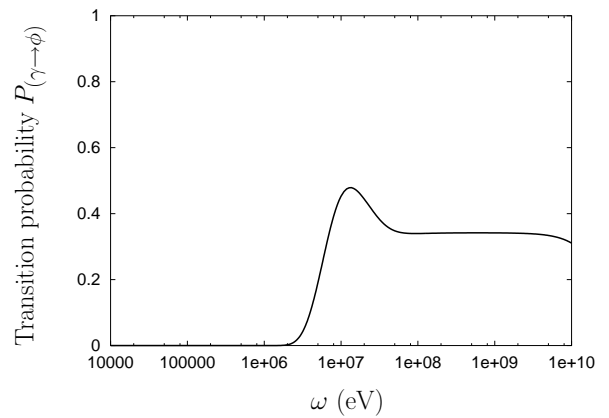}
\caption{\label{fig1}{Transition probability versus photon energy for a trajectory of the light beam with an impact parameter $\rho_0 =4 \cdot 10^3$ km.}}
\end{center}
\end{figure}

We see that photon-LPB conversion turns out to be important for $\omega > 10 \, {\rm MeV}$, namely in the $\gamma$-ray band. This is a remarkable conclusion, since pulsar $A$ in J0737-3039 is expected to be a $\gamma$-ray source. Furthermore, $\gamma$-ray photons propagate totally unimpeded in the magnetosphere of pulsar $B$, and so we do not have to bother about further potential sources of beam attenuation (photo-pair production is totally negligible). 

As a matter of fact, the temporal behaviour of the considered effect is best expressed in terms of the total {\it transmission} $A = 1 - P(\gamma \to \phi)$ of the beam after propagation in the magnetosphere of pulsar $B$. We plot $A$ versus time in Fig. 2, as pulsar $A$ moves in its nearly edge-on orbit. 

\begin{figure}[h]
\begin{center}
\includegraphics[width=8cm]{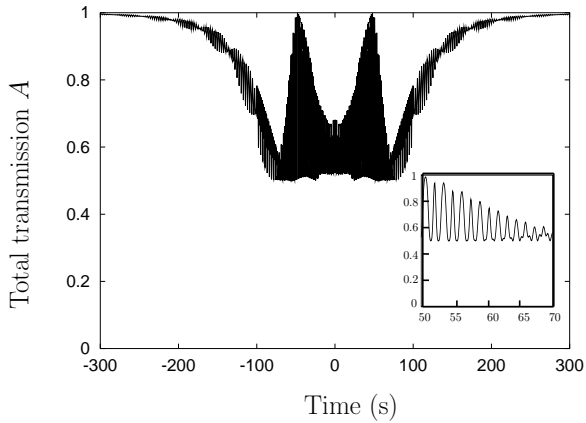}
\caption{\label{fig2}{Total transmisson of the gamma photon beam emitted by
    pulsar $A$ versus
    time. Inset shows the modulation mainly due to the rotation of the
    magnetic dipole moment of pulsar $B$.}}
\end{center}
\end{figure}

Our numerical simulation predicts a strong attenuation of the photon beam up to $50\,\%$ with a time duration of about $200$ s. As it is clear from Fig. 2, this effect has three different temporal structures. The broad minimum -- from $-200$ s to $+200$ s -- evidently corresponds to the transit of pulsar $A$ behind pulsar $B$. The tens-of-seconds, symmetric peaks are due to photon-LPB oscillations, depending on the actual path through the interaction region with pulsar $B$. Finally, the highest frequency modulation -- shown in inset -- is due to the rotation of the magnetic dipole moment of pulsar $B$. 
 
Finally, we show in Fig. 3 the region of the $m-M$ parameter plane excluded by the no detection of an attenuation $A$ at the $10 \, \%$ level. Such an attenuation is achieved by a 100 photon count during the total integration time. For a two weeks observation time, this corresponds to a flux from pulsar $A$ of about $2 \cdot 10^{-7}$ photons/cm$^2$/s, which is a reasonable flux according to previous observations of several pulsars. As a matter of fact, the GLAST sensitivity curves allow for a much weaker minimum detectable flux, down to $1 \cdot 10^{-10}$ 
photons/cm$^2$/s. 

Thus, we conclude that the photon-LPB conversion mechanism in the double pulsar J0737-3039 really provides a cross-check for the recent PVLAS claim about the existence of a new 
LPB. We stress that our result would remain practically unchanged even if we were considering a light {\it scalar} boson.
\begin{figure}[h]
\begin{center}
\includegraphics[width=8cm]{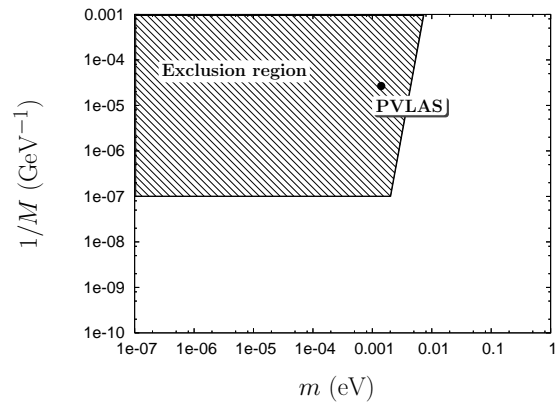}
\caption{\label{fig3}{Exclusion region in the case that the existence of the attenuation is excluded at $10\,\%$ level.}}
\end{center}
\end{figure}

\section{ACKNOWLEDGMENTS}

We thank our collaborators Nanni Bignami and Carlo Rizzo, with whom the work reported in 
this Proceeding was done. One of us (M. R.) would also like to thank professor Gianluigi Fogli for his kind invitation to talk at this beautiful workshop.


\begin{thebibliography}{9}
\bibitem{axion} J. H. Kim, Phys. Rep. {\bf 150} (1987) 1. H. Y. Cheng, Phys. Rep. 
                {\bf 158} (1988) 1.

\bibitem{cgn} S. L. Cheng, C. Q. Geng and W. T. Ni, Phys. Rev. D {\bf 94} (1995) 3132.

\bibitem{masso1} E. Masso and R. Toldra, Phys. Rev. D {\bf 52} (1995) 1755. E. Masso and R. Toldra, Phys. Rev. D {\bf 55} (1997) 7967.  

\bibitem{Heisenberg} W. Heisenberg and H. Euler, Z. Phys. {\bf 98} (1936) 714. J. Schwinger, Phys. Rev. {\bf 82} (1951) 664. W. Dittrich and H. Gies, {\it Probing the Quantum Vacuum} (Springer, Berlin, 2000).

\bibitem{Maiani} L. Maiani, R. Petronzio and E. Zavattini, Phys. Lett. B {\bf 175} (1986) 359.

\bibitem{pvlas} E. Zavattini {\it et al.}, Phys. Rev. Lett. {\bf 96} (2006) 110406.

\bibitem{cameron} R. Cameron {\it et al.}, Phys. Rev. D {\bf 47} (1993) 3707.

\bibitem{pvlas2} E. Zavattini {\it et al.}, in {\it Eleventh International Workshop on Neutrino Telescopes}, ed. by M. Baldo Ceolin (Papergraf, Padua, 2005). 

\bibitem{pvlas3} Private communication from the PVLAS collaboration.

\bibitem{Raffelt1996} G. G. Raffelt, {\it Stars as Laboratories for Fundamental Physics} (University of Chicago Press, Chicago, 1996). 

\bibitem{cast} K. Zioutas {\it et al.}, Phys. Rev. Lett. {\bf 94} (2005) 121301.

\bibitem{masso2005} E. Masso and J. Redondo, J. Cosmol. Astropart. Phys. {\bf 09} (2005) 015. P. Jain and S. Mandal, astro-ph/0512155 (2005).

\bibitem{masso22005} E. Masso and J. Redondo, Phys. Rev. Lett. {\bf 97} (2006) 151802. T. Fukuyama and T. Kikuchi, hep-ph/0608228 (2006). S. A. Abel, J. Jaeckel, V. V. Khoze and A. Ringwald, hep-ph/0608248 (2006). R. N. Mohapatra and S. Nasri, hep-ph/0610068 (2006).

\bibitem{drrb} A. Dupays, C. Rizzo, M. Roncadelli and G. F. Bignami, Phys. Rev. Lett. {\bf 95} (2005) 211302.

\bibitem{burgay} M. Burgay {\it et al.}, Nature (London) {\bf 426}, 531 (2003).

\bibitem{coles} W. A. Coles {\it et al.}, Astrophys. J. {\bf 623} (2005) 392.

\bibitem{Raffelt and Stodolsky} G. Raffelt and L. Stodolsky, Phys. Rev. D {\bf 37} (1988) 1237.



\end{thebibliography}
\end{document}